Doughnut  

\documentclass[10pt,letterpaper]{article}
\usepackage{opex3}
\usepackage{amsmath}
\usepackage{bm}
\usepackage{amscd}
\usepackage{amssymb}
\usepackage{graphicx}
\usepackage{epsfig}
\usepackage{mathrsfs}
\usepackage{amsfonts}
\usepackage{dcolumn}

\begin{document}

\title{Classical and quantum properties of cylindrically polarized states of light}

\author{Annemarie Holleczek$^{*1, 2}$, Andrea Aiello$^{1,2}$, Christian Gabriel$^{1,2}$, \\Christoph Marquardt$^{1,2}$, Gerd Leuchs$^{1,2}$}

\address{$^1$Max Planck Institute for the Science of Light, G\"{u}nter-Scharowsky-Str. 1/Bau 24, 91058 Erlangen, Germany}
\address{$^2$Institute for Optics, Information and Photonics, University of Erlangen-Nuremberg,\\ Staudtstr. 7/B2, 91058 Erlangen, Germany}

\email{annemarie.holleczek@mpl.mpg.de} 



\begin{abstract}
We investigate theoretical properties of beams of light with non-uniform polarization patterns. Specifically, we determine all possible configurations of cylindrically polarized modes (CPMs) of the electro-magnetic field, calculate their total angular momentum and highlight the subtleties of their structure. Furthermore, a hybrid spatio-polarization description for such modes is introduced and developed. In particular, two independent Poincar\'e spheres have been introduced to represent simultaneously the polarization and spatial degree of freedom of CPMs. Possible mode-to-mode transformations accomplishable with the help of conventional polarization and spatial phase retarders are shown within this representation. Moreover, the importance of these CPMs in the quantum optics domain due to their classical features is highlighted. 
\end{abstract}




\section{Introduction}

States of light with spatially inhomogeneous polarization structure, such as the radially or azimuthally polarized vector beams \cite{Siegman, RadAzi1, Sheppard2000}, have lately attracted major attention both in the theoretical and experimental area of research in optics \cite{Maurer2007}. As devices for their generation have become commercially available, their application in optical science and engineering are widespread -- ranging from classical to quantum optics experiments. In classical optics, when strongly focussed, these complex field distributions exhibit spatially separated longitudinal and transversal electric and magnetic field components depending on the initial state of polarization \cite{Quabis20001, classical}. This means, these vector beams can be sharply focussed in the center of the optical axis and therefore find applications in lithography, confocal microscopy \cite{Huse2001}, optical trapping and tweezing \cite{Sondermann2007} as well as material processing \cite{Meier2006}. Moreover, one can gain detailed insight into nanoscale physics due to the intrinsic strong polarization dependence of these beams \cite{Banzer2010}. In the quantum regime, specially designed spatio-polarization modes can increase the coupling to single ions \cite{Sondermann2007}. Additionally, the related orbital angular momentum states help to investigate entanglement \cite{Lassen2009}. Very recently it has been shown that by quadrature squeezing an azimuthally polarized optical beam, one can generate quantum states exhibiting \emph{hybrid} entanglement between the spatial and the polarization degrees of freedom \cite{Gabriel2010, Barreiro2010}. 

In this work we aim at establishing a proper theoretical framework for the classical optics description of cylindrically polarized modes (CPMs) of the electro-magnetic field. We cover the subtleties in terms of their description, highlight their symmetry properties and determine their total angular momentum. Finally, we show a suitable way of constructing a reasonable tool of displaying these modes on a defined pair of Poincar\'e spheres. In particular, in Sec. 1, we illustrate the rotation principles underlying the intrinsic \emph{cylindrical} polarization of the investigated beams. We introduce all possible configurations of these modes and calculate the Schmidt rank, as this is a convincing way to determine the potential inseparability of these modes. We complete the theoretical investigation by deducing the total angular momentum of these CPMs in Sec. 3. In Sec. 4, we introduce our Poincar\'e sphere representation for the hybrid polarization/spatial degrees of freedom characterizing CPMs. In Sec. 5, we discuss the manipulations of these states of light on these hybrid Poincar\'e spheres with the help of conventional phase retarders. Furthermore, in Sec. 6, we extend our formalism to the quantum domain and illustrate the importance of these modes in quantum experiments. Finally, in the Appendix, we furnish the reader with the mathematical tools which are essential to fully cover the topics of the main text.

\section{Classical description of light beams with cylindrical polarization}

\subsection{Introduction to rotation principles}

To mathematically introduce the radially and azimuthally polarized vector fields, let us consider a well collimated, monochromatic light beam with the wavelength $\lambda = 2\pi/k$ propagating along the $z$-axis. The corresponding electric field can be written as
\begin{equation}\label{Efeld}
\mathbf{E}(\mathbf{r},t) = \frac{1}{2}\bigg ( \mathbf{u}(\mathbf{r})e^{-i\chi} + \mathbf{u}^{*}(\mathbf{r})e^{i\chi}\bigg ),
\end{equation}
where $\chi = k(ct-z) - \pi/2$ \cite{Loudon} and $\mathbf{r} = x\hat{\mathbf{x}} +y\hat{\mathbf{y}} + z\hat{\mathbf{z}} $. Furthermore, $\mathbf{u}(\mathbf{r})$ is a paraxial mode function of the form 
\begin{equation}
\mathbf{u}(\mathbf{r}) = f_1(\mathbf{r})\hat{\mathbf{x}}+  f_2(\mathbf{r})\hat{\mathbf{y}}.
\label{mode_function}
\end{equation}
The unit vectors $\hat{\mathbf{x}}$ and $\hat{\mathbf{y}}$ indicate the polarization in the $x$ and $y$ directions and $f_1(\mathbf{r})$ and $f_2(\mathbf{r})$ are two independent solutions of the paraxial wave equation:
\begin{equation}
\frac{\partial ^2 f_{1,2}}{\partial x^2} +\frac{\partial ^2 f_{1,2}}{\partial y^2}  + 2ik\frac{\partial f_{1,2}}{\partial z} =0.
\label{paraxial_wave_equation}
\end{equation}

\begin{figure}
\begin{center}
\includegraphics[angle=0,width=6truecm]{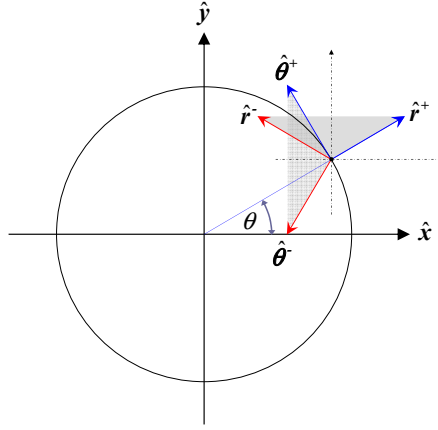}
\caption{\label{fig:1} Illustrating the {two} sets of pairs of {co-} and {counter-}rotating unit vectors $\{\hat{\mathbf{r}}^+(\theta),\hat{\bm{\theta}}^+(\theta)\}$ and $\{ \hat{\mathbf{r}}^-(\theta),\hat{\bm{\theta}}^-(\theta) \}$, respectively. The vector $\hat{\mathbf{r}}^-(\theta)$ is the mirror-image with respect to the vertical axis $y$
 of $\hat{\mathbf{r}}^+(\theta)$, while $\hat{\bm{\theta}}^-(\theta)$ is the mirror-image with respect to the horizontal axis $x$ of $\hat{\bm{\theta}}^+(\theta)$.}
\label{Fig1}
\end{center}
\end{figure}

In our investigation we are not interested in the beam propagation, and therefore we shall consider the longitudinal coordinate $z$ as fixed to a value which will be omitted in subsequent formulas. Therefore, we define the \emph{transverse} position vector $\mathbf{x}$ as:
\begin{equation}
\mathbf{x} = x \hat{\mathbf{x}} + y \hat{\mathbf{y}} \label{Azi35}.
\end{equation}
Given a \emph{global} Cartesian reference frame $\{ \hat{\mathbf{x}}, \hat{\mathbf{y}} \}$ on the $xy$-plane, it is always possible to define a \emph{local} (depending on the position vector $\mathbf{x}$) polar reference frame $\{\hat{\mathbf{r}}(\theta), \hat{\bm{\theta}}(\theta)\}$ with $x = r \cos \theta, \, y = r \sin \theta$, where $r \in [0, \infty[$ and $\theta \in [0, 2 \pi[$. The two coordinate systems are related through the following  relation:
\begin{subequations}\label{Azi70}
\begin{align}
\hat{\mathbf{r}}(\theta) =& \,   \cos \theta \,  \hat{\mathbf{x}} +\sin \theta \hat{\mathbf{y}}, \\
\hat{\bm{\theta}}(\theta) =& \, -  \sin \theta \,  \hat{\mathbf{x}} +\cos \theta \hat{\mathbf{y}}.
\end{align}
\end{subequations}
Since, by definition, in the  $\{ \hat{\mathbf{x}}, \hat{\mathbf{y}} \}$ basis one has
\begin{align}
\hat{\mathbf{x}} = \left(
                \begin{array}{c}
                  1 \\
                  0 \\
                \end{array}
              \right), \qquad \hat{\mathbf{y}} = \left(
                \begin{array}{c}
                  0 \\
                  1 \\
                \end{array}
              \right),
\end{align}
then from Eq. (\ref{Azi70}) it follows that
\begin{align} \label{Azi75}
    \hat{\mathbf{r}}(\theta) = R(\theta)\hat{\mathbf{x}}, \qquad         \hat{\bm{\theta}}(\theta)  = R(\theta)\hat{\mathbf{y}},
\end{align}
where $R(\theta)$ denotes the $2 \times 2$ rotation matrix which, in the $\{ \hat{\mathbf{x}}, \hat{\mathbf{y}} \}$ basis, can be written as:
\begin{align}
R(\theta) = \left(
                \begin{array}{cc}
                  \cos \theta & -\sin \theta \\
                  \sin \theta & \cos \theta \\
                \end{array}
              \right). \label{Azi50}
\end{align}
Since rotation operations are additive and commute in two-dimensional spaces,  from Eq. (\ref{Azi75}) it can be deduced that for any $\varphi \in \mathbb{R}$ the following holds:
\begin{align} \label{Azi85}
    R(\varphi)\hat{\mathbf{r}}(\theta) = R(\theta + \varphi)\hat{\mathbf{x}} = \hat{\mathbf{r}}(\theta + \varphi), \\     
    R(\varphi)\hat{\bm{\theta}}(\theta)  = R(\theta+ \varphi)\hat{\mathbf{y}}=\hat{\bm{\theta}}(\theta + \varphi).
\end{align}
These equations shows that $\hat{\mathbf{r}}(\theta)$ and $\hat{\bm{\theta}}(\theta)$ co-rotate with respect to the applied ``external'' rotation $R(\varphi)$. More generally, it is not difficult to prove that Eqs. (\ref{Azi75}) may be extended to  \emph{two} sets of pairs of \emph{co-} and \emph{counter-}rotating unit vectors $\{\hat{\mathbf{r}}^+(\theta),\hat{\bm{\theta}}^+(\theta)\}$ and $\{ \hat{\mathbf{r}}^-(\theta),\hat{\bm{\theta}}^-(\theta) \}$, respectively, defined as
\begin{align} \label{Azi75b}
    \hat{\mathbf{r}}^\pm(\theta) = \pm R(\pm \theta)\hat{\mathbf{x}}, \qquad         \hat{\bm{\theta}}^\pm(\theta)  = \pm R(\pm \theta)\hat{\mathbf{y}},
\end{align}
where now
\begin{align} \label{Azi85b}
    R(\varphi)\hat{\mathbf{r}}^\pm(\theta) =  \hat{\mathbf{r}}^{\pm}(\theta \pm \varphi), \quad   R(\varphi)\hat{\bm{\theta}}^\pm(\theta)  = \hat{\bm{\theta}}^{\pm}(\theta \pm \varphi).
\end{align}
The two sets of pairs of unit vectors are illustrated in Fig. \ref{Fig1}. From these equations it is clear that any either {co-rotating } $\mathbf{u}^+(r,\theta)$ or  {counter-}rotating   $\mathbf{u}^-(r,\theta)$ vector field may be written as:

\begin{align}
\mathbf{u}^\pm(r,\theta) = \hat{\mathbf{r}}^\pm(\theta) \, a(r) +   \hat{\bm{\theta}}^\pm(\theta) \,  b(r) , 
\label{Azi60}
\end{align}
where $a(r)$ and $b(r)$ are arbitrary functions independent of $\theta$ but, although omitted, still dependent of $z$. By definition, such fields
satisfy the following condition:
\begin{align}
\mathbf{u}^\pm(r,\theta \pm \varphi) = R(\varphi)\mathbf{u}^\pm(r,\theta) , 
\label{rot}
\end{align}
where with $R(\varphi)\mathbf{u}^\pm(r,\theta)$ we had indicated the ordinary matrix-times-vector
product, namely:
\begin{align}
R(\varphi)\mathbf{u}^\pm(r,\theta) = \left(
                \begin{array}{cc}
                  \cos \varphi & -\sin \varphi \\
                  \sin \varphi & \cos \varphi \\
                \end{array}
              \right) \left(
                \begin{array}{c}
                \hat{\mathbf{x}} \cdot \mathbf{u}^\pm\\
                \hat{\mathbf{y}} \cdot \mathbf{u}^\pm \\
                \end{array}
              \right). \label{Azi55}
\end{align}

Eq. (\ref{rot}) has some interesting consequences: First, we note that, since $x = r \cos \theta$ and $y = r \sin \theta$, we can rewrite $x = x(r, \theta)$ and $y = y(r , \theta)$, having assumed $r,\theta$ as independent variables. Then, by definition,
%
%
%
\begin{align}
x(r, \theta \pm \varphi) =  & \; r \cos (\theta \pm \varphi) = r (\cos \theta \cos \varphi \mp \sin \theta \sin \varphi), \label{Azi100}
\end{align}
and
\begin{align}
y(r, \theta \pm \varphi) =  & \; r \sin (\theta \pm \varphi) = r (\sin \theta \cos \varphi \pm \cos \theta \sin \varphi). \label{Azi110}
\end{align}
These two equations can be rewritten in compact vector notation as:
\begin{align}
\mathbf{x}(r, \theta \pm \varphi) = R(\pm \varphi) \mathbf{x}(r, \theta) . \label{Azi120}
\end{align}
Now, since the argument of the functions $\mathbf{u}^\pm(r,\theta)=\mathbf{u}^\pm(\mathbf{x})$ on the left side of Eq.  (\ref{rot}) is exactly $\mathbf{x}(r, \theta \pm \varphi)$, we can use the result above  to rewrite Eq. (\ref{rot}) as: %
\begin{align}
\mathbf{u}^\pm(R(\pm \varphi) \mathbf{x}) = R(\varphi)\mathbf{u}^\pm(\mathbf{x}) , 
\label{Azi130}
\end{align}
or, equivalently,
\begin{subequations}\label{Azi140}
\begin{align}
\mathbf{u}^+( \mathbf{x}) = & \; R(\varphi)\mathbf{u}^+(R(- \varphi)\mathbf{x}) , \label{Azi140a} \\
\mathbf{u}^-( \mathbf{x}) = & \; R(\varphi)\mathbf{u}^-(R(\varphi)\mathbf{x}). \label{Azi140b}
\end{align}
\end{subequations}
At this point, one should remember that any $2$-dimensional vector field $\mathbf{V}(x,y)$ transforms under a \emph{global}
counterclockwise active rotation by an angle $\varphi$ into the new field   $\mathbf{W}(x,y)$ whose functional expression is determined according to the following rule:
\begin{align}
\mathbf{W}( \mathbf{x}) = R(\varphi)\mathbf{V}(R(- \varphi) \mathbf{x}). \label{Azi145}
\end{align}
Thus Eq. (\ref{Azi140a}) coincides \emph{exactly} with the definition of a vector field which is invariant with respect to a \emph{global} rotation by an angle $\varphi$. This means that co-rotating beams have a polarization pattern \emph{invariant} with respect to global rotations. This result is intuitively obvious for radial and azimuthally polarized beams.
Conversely, the corresponding Eq. (\ref{Azi140b}) for the counter-rotating beams has a less straightforward geometric interpretation. For this case, one can show that  Eq. (\ref{Azi140b}) can be recast in the form of a global rotation as:
\begin{align}
R(2 \varphi)\mathbf{u}^-( \mathbf{x}) = & \; R(\varphi)\mathbf{u}^-(R(-\varphi)\mathbf{x}). \label{Azi147}
\end{align}
This equation simply shows that a global rotation by an angle $\varphi$ upon the counter-rotating modes is equivalent to a \emph{local} rotation by an angle $2 \varphi$ on the same mode.
\subsection{Deduction of cylindrically polarized modes}
As we are now knowledgeable about all rotation properties which the cylindrically polarized optical beams have to obey to, we want to explicitly derive all possible cylindrical configurations of these mode functions which fulfill Eq. (\ref{rot}) taken with the ``$+$'' sign. However, it will be eventually shown that these coincide with the well-known radially and azimuthally polarized beams. 

\begin{figure}
\includegraphics[width=12cm]{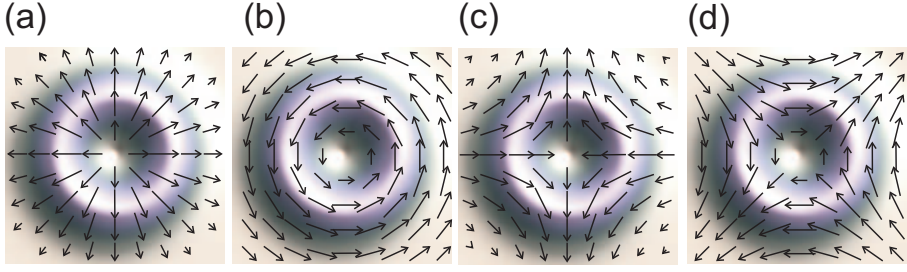}
\caption{Complex polarization patterns of (a) $\mathbf{u}^+_R$, (b) $\mathbf{u}^+_A$, (c) $\mathbf{u}^-_R$, (d) $\mathbf{u}^-_A$, underlayed with the doughnut shaped intensity distribution.}
\label{fig:basis}
\end{figure}

After inverting Eq. (\ref{Azi70})  and inserting the result into Eq. (\ref{mode_function}), we obtain as a paraxial mode function
\begin{equation}
\mathbf{u}(r, \theta) = \hat{\mathbf{r}}(\cos\theta f_1 + \sin\theta f_2) +\hat{\bm{\theta}}(-\sin\theta f_1 + \cos\theta f_2).
\label{mode_function_2}
\end{equation}
By comparing Eq. (\ref{mode_function_2}) with Eq. (\ref{Azi60}), one obtains at once:
\begin{subequations}
\begin{align} \label{f1f2a}
f_1(\mathbf r) &= a(r) \cos\theta - b(r)\sin\theta,\\
 \label{f1f2b}
f_2(\mathbf r) &= a(r) \sin\theta + b(r)\cos\theta.
\end{align}
\end{subequations}
Since $x= r\cos{\theta}$ and $y= r\sin{\theta}$, Eqs. (\ref{f1f2a}) and  (\ref{f1f2b}) can be rewritten as:
\begin{subequations}
\begin{align} \label{f1f2_2a}
f_1(\mathbf r) &= x\alpha(r)  - y\beta(r),\\
f_2(\mathbf r) &=y \alpha(r) + x\beta(r),
\label{f1f2_2b}
\end{align}
\end{subequations}
where $\alpha(r)$ and $\beta(r)$ can be determined by keeping in mind that $f_1$ and $f_2$ need to satisfy the paraxial equation (\ref{paraxial_wave_equation}). In particular, we want to express the functions $f_1(\mathbf r)$ and $f_2(\mathbf r)$ in terms of the first-order Hermite-Gauss \cite{Padgett1999} solutions of the paraxial wave equation which can be -- apart from an irrelevant normalization factor $\sqrt{8/\pi}$ -- written at $z=0$, as
 
\begin{subequations}
\begin{align} \label{HermiteGauss_a}
\psi_{10}(\mathbf{x}) = x e^{-(x^2+y^2)/w_0^2}/w_0,\\
\psi_{01}(\mathbf{x}) = ye^{-(x^2+y^2)/w_0^2}/w_0,
\label{HermiteGauss_b}
\end{align}
\end{subequations}
where $w_0$ is the beam waist. Thus, from Eqs. (\ref{f1f2_2a}) and (\ref{f1f2_2b}) and the fact that $x\propto \psi_{10}$ and $y\propto \psi_{01}$, it follows that:
\begin{subequations}
\begin{align}
f_1(\mathbf r) = \frac{1}{\sqrt{2}} (A\psi_{10} - B\psi_{01}), \\
f_2(\mathbf r) = \frac{1}{\sqrt{2}} (B\psi_{10} + A\psi_{01}),
\label{Azi140}
\end{align}
\end{subequations}
where $A,\,B$ are arbitrary numerical constants. In general, with $\psi _{nm} = \psi_{nm}(x,y,z)$ ($n, m \in \mathbb N_0$) we denote the Hermite-Gauss solutions of the paraxial wave equation of the order $N = n+m$ \cite{Padgett1999}. Consequently, all possible solutions which fulfill the global law of Eq. (\ref{rot}) with the ``$+$'' sign, obviously have the form of:
\begin{equation}
\mathbf{u} =  \frac{1}{\sqrt{2}} (A\psi_{10} - B\psi_{01})\hat{\mathbf{x}} +\frac{1}{\sqrt{2}} (B\psi_{10} + A\psi_{01})\hat{\mathbf{y}}. 
\label{eq:allgemein}
\end{equation}
By choosing either $\{A=1, B=0\}$ or $\{A=0, B=1\}$ and inserting this in Eq. (\ref{eq:allgemein}), one obtains after normalization:
\begin{subequations}
\begin{align} \label{eq:uR+uA+a}
\hat{\mathbf{u}}_R^{+} &=  \frac{1}{\sqrt{2}}( \psi _{10}\hat{\mathbf{x}}+  \psi_{01}\hat{\mathbf{y}}),\\
 \hat{\mathbf{u}}_A^{+} &= \frac{1}{\sqrt{2}}( - \psi _{01}\hat{\mathbf{x}}+  \psi_{10}\hat{\mathbf{y}}).
 \label{eq:uR+uA+b}
\end{align}
\end{subequations}
The same steps that lead us to Eqs. (\ref{eq:uR+uA+a}) and (\ref{eq:uR+uA+b}) could be repeated for the counter-rotating modes. However, it is more instructive to note that, from a visual inspection, one can immediately deduce that $\hat{\mathbf{u}}_R^{+}$ and $\hat{\mathbf{u}}_A^{+}$ live in a four dimensional space spanned by the basis formed by the Cartesian product of the $\{\psi_{10}, \psi_{01}\}$ mode bases and the polarization vectors $\{\hat{\mathbf{x}}, \hat{\mathbf{y}} \}$:
\begin{equation}
 \{  \psi_{10}, \psi_{01}  \}\otimes\{ \mathbf{\hat x}, \mathbf{\hat y}\}=\{  \psi_{10}\mathbf{\hat x},  \psi_{10}\mathbf{\hat y}, \psi_{01}\mathbf{\hat x}, \psi_{01}\mathbf{\hat y} \}.
\label{eq:space}
\end{equation} 
However, the two vectors $\{\mathbf{u}_R^+,\mathbf{u}_A^+\}$ only form a two dimensional basis and consequently two more vectors have to be added to span the whole space. By applying a unitary transformation to such a basis, we can obtain a new set of basis vectors, namely $\{\mathbf{u}_R^-,\mathbf{u}_A^-\}$, where:
\begin{subequations}
\begin{align} \label{eq:uR-uA-a}
\hat{\mathbf{u}}_R^{-} &=  \frac{1}{\sqrt{2}}(- \psi _{10}\hat{\mathbf{x}}+  \psi_{01}\hat{\mathbf{y}}),\\
 \hat{\mathbf{u}}_A^{-} &= \frac{1}{\sqrt{2}}(  \psi _{01}\hat{\mathbf{x}}+  \psi_{10}\hat{\mathbf{y}}).
 \label{eq:uR-uA-b}
\end{align}
\end{subequations}
The vectors $\{\hat{\mathbf{u}}_R^{+}, \hat{\mathbf{u}}_A^{+}, \hat{\mathbf{u}}_R^{-},  \hat{\mathbf{u}}_A^{-}  \}$, whose intensity and polarization profiles are illustrated in Fig. \ref{fig:basis}, form a complete four-dimensional basis, orthonormal with respect to the scalar product defined as:
\begin{align}
&\big (\mathbf{u} (\mathbf{x}),\mathbf{v}({\mathbf{x}})\big) = \sum_{i=1}^{2} \iint_{-\infty}^{\infty} \,u_{i}^*(\mathbf{x})v_{i}(\mathbf{x}) \,dxdy,
\end{align}
where $\mathbf{u}(\mathbf{x})$ and $\mathbf{v}(\mathbf{x})$ are two arbitrary vector functions.
It should be stressed that the basis vectors $\hat{\mathbf{u}}_R^{-} $ and $\hat{\mathbf{u}}_A^{-}$ do fulfill the rotation law of Eq. (\ref{rot}), but with the ``$-$'' sign:
\begin{equation}
\mathbf{u}^{-}(r, \theta - \alpha) = R(\alpha)\mathbf{u}^{-}(r, \theta).
\label{rotMinus}
\end{equation}
Furthermore, Eqs. (\ref{rot}) and (\ref{rotMinus}) imply that these two sets of modes do not mix under rotation. This is interesting, as many recent experiments show \cite{classical, Gabriel2010} that co- and counter-rotating modes do not occur simultaneously.
\subsection{Schmidt rank of radially and azimuthally polarized modes}
Another important property of CPMs can be characterized by the so-called Schmidt rank of the modes. The Schmidt rank \cite{Peres} is an intriguing quantity as it provides information about potential inseparability of these modes. In the following, we demonstrate how to determine this rank for the radially and azimuthally polarized optical beams. This can be similarly calculated for the counter-rotating modes.

If one has a look at the structure of these basis vectors, one can easily see that they are clearly \emph{non-separable}, in the sense that it is not possible to write any of these vector functions as the product of a uniformly polarized vector field times a scalar function. Such a ``degree of non-separability'' may be quantified by the Schmidt-rank of the function, remembering that all separable functions have a Schmidt rank of 1. Conversely, we are now going to prove that $\hat{\mathbf{u}}_R^{+}$ and $\hat{\mathbf{u}}_A^{+}$ admit a Schmidt decomposition with rank $2$:

Let $\hat{\mathbf{e}}_1$ and $\hat{\mathbf{e}}_2$ be two orthogonal unit vectors defined as $\hat{\mathbf{e}}_1 = \hat{\mathbf{x}}$ and $\hat{\mathbf{e}}_2 =  \hat{\mathbf{y}}$, with
\begin{equation*}
 \mathbf{e}_i^* \cdot \mathbf{e}_j  = \delta_{ij}, \qquad i,j\in\{1,2\},
\end{equation*}
and let $v_1$ and $v_2$ be two orthogonal functions defined as
\begin{equation}\label{SepScal}
v_1 = \psi_{10}(x,y,z), \qquad v_2 = \psi_{01}(x,y,z),
\end{equation}
where, because of the orthogonality of the Hermite-Gauss modes, one has:
\begin{equation*}
 \iint _{-\infty} ^{\infty}v_i^* v_j \,dx dy= \delta_{ij}, \qquad i,j\in\{1,2\}.
\end{equation*}
Then we can immediately write down, for example for the $\hat{\mathbf{u}}_R^+$ mode:
\begin{equation}\label{Sep3}
\hat{\mathbf{u}}_R^+ = \sum_{i=1}^2 (\lambda_i)^{1/2} \, \hat{\mathbf{e}}_i\, v_i,
\end{equation}
with $\lambda_1 = \lambda_2 = 1/2 $. The right side of Eq. (\ref{Sep3}) represents the Schmidt decomposition of the vector function $\hat{\mathbf{u}}_R^+$ with rank
\begin{equation}
K = {1}/{\displaystyle{\sum_{i=1}^2 \lambda_i^2}} = 2. 
\label{K}
\end{equation}

For the function $\hat{\mathbf{u}}_A^+$ the same demonstration holds but with $v_1$ and $v_2$  defined as
\begin{equation}\label{SepScalA}
v_1 = -\psi_{01}(x,y,z), \qquad v_2 = \psi_{10}(x,y,z),
\end{equation}
This completes the proof. 

More generally, the substitution of Eq. (\ref{eq:uR+uA+a}) or (\ref{eq:uR+uA+b}) respectively into Eq. (\ref{rot}) gives

\begin{align}\label{Azi240}
\hat{\mathbf{u}}_{AB}^{+} = & \;  \frac{1}{\sqrt{2}} \left[ \psi_{10} \left( A \hat{\mathbf{x}} + B  \hat{\mathbf{y}}\right) + \psi_{01} \left(-B\hat{\mathbf{x}} + A  \hat{\mathbf{y}} \right) \right] \nonumber \\
\equiv & \;   \frac{1}{\sqrt{2}}  \,\mathbf{f}_1  \psi_{10} +  \frac{1}{\sqrt{2}}  \, \mathbf{f}_2 \psi_{01},
\end{align}
where we have chosen $|A|^2 + |B|^2 =1$ to guarantee normalization: $(\hat{\mathbf{u}}_{AB}, \hat{\mathbf{u}}_{AB}) =1$. From Eq. (\ref{K}) it follows that $\hat{\mathbf{u}}_{AB}^{+}$ has Schmidt rank $2$ whenever 
\begin{equation}
\mathbf{f}_1^* \cdot  \mathbf{f}_2= A B^* - A^* B=0, 
\end{equation}
namely when $\text{Im}(A B^*) =0$. Conversely,  $\hat{\mathbf{u}}_{AB}^+$ becomes separable when $\mathbf{f}_2 = C \mathbf{f}_1$, namely when $-B = C A, \; A = C B$, where $C$ is a proportionality constant. The substitution of the latter equation in the previous one furnishes $-B = C^2 B$ which implies $C = \pm i$, namely $A = \pm i  B$. In this case it is easy to see that Eq. (\ref{K}) reduces to
\begin{align}\label{68}
\hat{\mathbf{u}}_{AB} ^{+} =\frac{A}{\sqrt{2}} \left( \hat{\mathbf{x}}\mp i  \hat{\mathbf{y}} \right)\left(  \psi_{10} \pm i \psi_{01} \right) ,
\end{align}
which means that the only separable cylindrically symmetric co-rotating modes are the circularly polarized Laguerre-Gauss modes with one unit ($|\ell| = 1$) of orbital angular momentum: $\phi^\text{LG}_{\pm1, 0} = (  \psi_{10} \pm i \psi_{01})/\sqrt{2}$. It should be noticed that it is only thanks to the opposite double sign in  the last two terms of Eq. (\ref{68}) that condition (\ref{rot}) is satisfied.
\section{Angular momentum of cylindrically polarized optical beams}

To complete our discussion about the properties of the ``$+$'' and ``$-$'' vector bases, we note that all four vector functions $\{ \mathbf{u}^+_R,\mathbf{u}^-_R,\mathbf{u}^+_A,\mathbf{u}^-_A \}$ have the total (orbital plus spin) angular momentum (TAM)  of zero which is deduced below. Following Berry \cite{Berry2009}, we can write the ``local'' (position-dependent) density of the \emph{linear momentum} carried by the paraxial mode $\mathbf{u}$ of the electromagnetic field as:
\begin{align}\label{AM30}
\mathbf{p}(\mathbf{r}) = & \; \text{Im} \big[ \mathbf{u}^* \cdot (\nabla_{\! \perp}) \mathbf{u} \big] +   \frac{1}{2} \text{Im} \big[ \nabla_{\! \perp} \times \left( \mathbf{u}^* \times \mathbf{u} \right) \big] \nonumber \\
\equiv & \; \mathbf{p}_\text{orb}(\mathbf{r}) + \mathbf{p}_\text{sp}(\mathbf{r}),
\end{align}
where $\nabla_{\! \perp} = \hat{\mathbf{x}}\partial_x + \hat{\mathbf{y}}\partial_y$ and $\mathbf{u}^* \cdot (\nabla_{\! \perp}) \mathbf{u} = \sum_{i} u^{*}_i( \nabla_{\! \perp} u_i) $. A prefactor of $c^2 \varepsilon_0/(2 \omega)$ has been omitted. The symbol ``$\times$'' denotes the ordinary cross product in $\mathbb{R}^3$. In Eq. (\ref{AM30}), the first term $\mathbf{p}_\text{orb}(\mathbf{r})$ is called the orbital Poynting current and  is independent of the polarization state of the beam. Conversely the second term, the so-called spin Poynting current $\mathbf{p}_\text{sp}(\mathbf{r})$, contains the local ``density of spin'' of the beam via the term $\text{Im}\left( \mathbf{u}^* \times \mathbf{u} \right)$ \cite{Berry2009}.
An explicit calculation from Eq. (\ref{AM30}) furnishes:
\begin{subequations}\label{AM40}
\begin{align}
 p_x(\mathbf{r}) = & \; \frac{1}{k}\text{Im} \left[ f_1^* (\partial_x f_1) +  f_2^*( \partial_x f_2) \right]  + \frac{1}{k}\text{Im}\left[ \partial_y ( f_1^* f_2 ) \right], \label{AM40x} \\
 p_y(\mathbf{r}) = & \; \frac{1}{k}\text{Im} \left[ f_1^* (\partial_y f_1) +  f_1^*( \partial_y f_2) \right]  - \frac{1}{k}\text{Im} \left[\partial_x ( f_1^* f_2 ) \right], \label{AM40y} \\
 p_z(\mathbf{r}) = & \; \left|f_1  \right|^2 + \left|f_2  \right|^2, \label{AM40z}
\end{align}
\end{subequations}
where $k$ indicates the wave vector $k = \omega/c$.
The power of a light beam as actually measured by a photo-detector in the laboratory equals the flow of the Poynting vector of the beam across the detector surface. Since the electromagnetic linear momentum density $\mathbf{p}$ is equal to $1/c^2$ times the Poynting vector,  it follows that a quantity experimentally observable  is $\mathbf{P} \cdot \hat{\mathbf{z}}$, where
\begin{align}\label{AM50}
\mathbf{P} = & \;\iint \mathbf{p}(\mathbf{r}) \, dx dy \nonumber \\
\equiv & \; \mathbf{P}_\text{orb} + \mathbf{P}_\text{sp},
\end{align}
is the linear momentum of the field per \emph{unit length}. Without loss of generality, we have assumed that the detector surface coincides with the $xy$-plane normal to the beam propagation axis $\hat{\mathbf{z}}$. The double integration in Eq. (\ref{AM50}) is extended to the whole $xy$-plane. This means that for any pair of functions $f_1$ and $f_2$, normalizable in the $xy$-plane, which drop to zero at infinity, the ``spin'' surface terms $\partial_y ( f_1^* f_2 )$  and $\partial_x ( f_1^* f_2 )$ on the right sides of Eq. (\ref{AM40x}) and Eq. (\ref{AM40y}) respectively, do not contribute to the measured  momentum. In other words one has $\mathbf{P}_\text{sp} =0$. It is worth noting that although $f_1$ and $f_2$ are both functions of $\mathbf{r}$, as opposed to $\mathbf{x}$   we can practically evaluate Eq. (\ref{AM50}) at $z=0$ because the quantity $\mathbf{P}$ is independent of $z$ \cite{AielloGSHEL}.

The \emph{total angular momentum density} is
\begin{align}\label{AM60}
\mathbf{j}(\mathbf{r}) = & \; \mathbf{r} \times \mathbf{p}(\mathbf{r}) \nonumber \\
= & \; \mathbf{r} \times \left[ \mathbf{p}_\text{orb}(\mathbf{r}) + \mathbf{p}_\text{sp}(\mathbf{r})\right] \nonumber \\
\equiv & \; \mathbf{l}(\mathbf{r}) + \mathbf{s} (\mathbf{r}) ,
\end{align}
where the \emph{orbital angular momentum density} is 
\begin{subequations}\label{AM70}
\begin{align}
 l_{x}(\mathbf{r}) = & \; y (|f_1 |^2 + |f_2 |^2)- \frac{z}{k}\text{Im} \left( f_1^* \partial_y f_1 +  f_2^* \partial_y f_2 \right), \label{AM70ox} \\
 l_{y}(\mathbf{r}) = & \; -x  (|f_1 |^2 + |f_2 |^2)+ \frac{z}{k}\text{Im} \left( f_1^* \partial_x f_1 +  f_2^* \partial_x f_2 \right), \label{AM70oy} \\
 l_{z}(\mathbf{r}) = & \; \text{Im} \left[x \!  \left( f_1^* \partial_y f_1 +  f_2^*\partial_y f_2 \right) - y \!  \left( f_1^* \partial_x f_1 +  f_2^* \partial_x f_2 \right) \right], \label{AM70oz}
\end{align}
\end{subequations}
and the spin angular momentum density is 
\begin{subequations}\label{AM80}
\begin{align}
 s_{x}(\mathbf{r}) = & \; -\frac{z}{k}\text{Im} \left[  \partial_x ( f_1^* f_2 ) \right], \label{AM80sx} \\
 s_{y}(\mathbf{r}) = & \; \frac{z}{k}\text{Im} \left[  \partial_y ( f_1^* f_2 ) \right], \label{AM80sy} \\
 s_{z}(\mathbf{r}) = & \; -\frac{1}{k}\text{Im} \left[x  \partial_x ( f_1^* f_2 ) + y  \partial_y ( f_1^* f_2 ) \right]. \label{AM80sz}
\end{align}
\end{subequations}
We proceed as in the linear momentum case and we write the angular momentum of the field per \emph{unit length} as
\begin{align}\label{AM90}
\mathbf{J} = & \;\iint \mathbf{j}(\mathbf{r}) \, dx dy \nonumber \\
\equiv & \; \mathbf{L} + \mathbf{S},
\end{align}
were, similarly to $\mathbf{P}$,   also $\mathbf{L}$ and  $\mathbf{S}$ are independent of $z$ \cite{AielloGSHEL}.
Again, for any pair of functions $f_1(\mathbf{r})$ and $f_2(\mathbf{r})$ 
that vanish sufficiently fast for $|x y| \rightarrow \infty$, integration by
parts leads to $S_x = 0 =S_y$, and to
\begin{align}\label{AM100}
\iint \! \!   \big[ x  \partial_x ( f_1^* f_2 ) + y  \partial_y ( f_1^* f_2 ) \big]  dx dy  = - 2 \!  \iint  \! \!  f_1^* f_2 \, dx dy.
\end{align}
From this  relation it follows straightforwardly  that
\begin{align}\label{AM110}
S_z = - \frac{i}{k} \iint \big( f_1^* f_2 - f_1 f_2^*\big)  \, dx dy.
\end{align}
Eq. (\ref{AM110}) clearly reproduces the well-know expression of the helicity $S_z  = i (\alpha \beta^* - \alpha^* \beta)/k$ of a  uniformly-polarized beam when one chooses $f_1 = \alpha \psi(\mathbf{r})$ and $f_2 = \beta \psi(\mathbf{r})$, where $ \psi(\mathbf{r})$ is a normalizable function in the $xy$-plane and $\alpha, \beta$ a pair of complex number such that $|\alpha|^2 + |\beta|^2=1$.

At this point, we can easily evaluate $\mathbf{J}$ for the set of our four cylindrically polarized modes without performing any actual calculations. In fact, since $\mathbf{J}$ is independent of $z$, we can evaluate it at $z=0$ where both $\psi_{10}(x,y,z=0)$ and $\psi_{01}(x,y,z=0)$ are  \emph{real-valued} functions. Then from Eqs. (\ref{AM70}-\ref{AM80}) it follows at once that $\mathbf{S} = \mathbf{0}$ and that
\begin{subequations}\label{AM120}
\begin{align}
 L_{x} = & \; \iint y (|f_1 |^2 + |f_2 |^2) \, dx dy, \label{AM120ox} \\
 L_{y} = & \; - \iint x  (|f_1 |^2 + |f_2 |^2) \, dx dy, \label{AM120oy} \\
 L_{z} = & \; 0.  \label{AM120oz}
\end{align}
\end{subequations}
However, it is easy to see from Eqs. (\ref{HermiteGauss_a}) and (\ref{HermiteGauss_b}) that by definition $|\psi_{10}(x,y,0)|^2$ and $|\psi_{01}(x,y,0)|^2$ are even functions with respect to both $x$ and $y$ coordinates. This automatically implies, because of symmetry, that $L_x = 0 = L_y$.

We have thus demonstrated that the TAM of our modes is strictly zero. Alternatively, the same result could have  been found in a more intuitive but less formal way by writing the modes  $\{ \mathbf{u}^+_R,\mathbf{u}^-_R,\mathbf{u}^+_A,\mathbf{u}^-_A \}$ in the left/right circular polarization basis and in the $\phi ^{LG}_{\pm1,0}$ Laguerre-Gauss  spatial basis. In this case, both the polarization and spatial modes carry a unit of angular momentum which may be summed or subtracted from each other, thus leading to a straightforward physical interpretation.

\section{Hybrid Poincar\'e sphere representation}

\begin{figure}
\centering
\includegraphics[width=4cm]{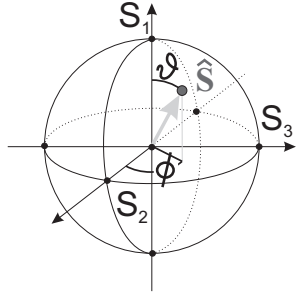}
\caption{Poincar\'e sphere representation for an arbitrary two-dimensional system. Here, the radius $r$ of the sphere is fixed to one, and spherical angles $\{\vartheta, \phi\}$ are related to the Stokes parameters via the relation: 
	$\{\cos\vartheta,\sin\vartheta \cos\phi, \sin\vartheta\sin\phi\} = \{S_1/S_0, S_2/S_0,S_3/S_0\}$, where $S_0$ denotes the total intensity of the beam.}
\label{fig:poincare}
\end{figure}

\begin{figure}[!ht]
\centering
\includegraphics[width=\textwidth]{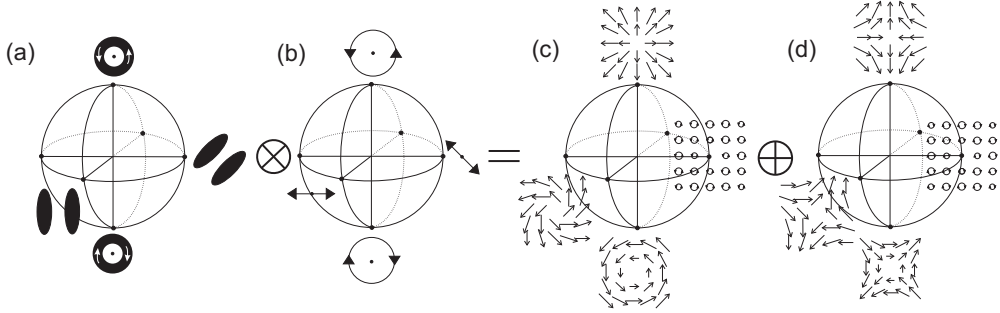}
\caption{Fundamental idea behind the new Poincar\'e sphere representation: Combining the SMPS (a) and the PPS (b) to the hybrid Poincar\'e spheres (HPSs) (c) and (d).} 
\label{fig:kombination}
\end{figure}

In general, fully polarized optical beams, similarly to quantum pure states of two-level systems \cite{Jauch}, can be described by points on a sphere of unit radius \cite{AzzamBook}, the so-called polarization Poincar\'e sphere [PPS, Fig. \ref{fig:kombination} (b)]. The North and South Poles of this sphere represent left-and right-handed polarization states that correspond to $\pm1$ \emph{spin} angular momentum eigenstates, respectively. Similarly, the spatial-mode Poincar\'e sphere [SMPS, Fig. \ref{fig:kombination} (a)] \cite{Padgett1999} sticks to the same superposition principle but in terms of the spatial distribution of the optical beams. In this case, the North and the South Pole of the sphere represent optical beams in the Laguerre-Gaussian modes $\phi_{\pm 1,0}^{LG}$, which correspond to $\pm1$ \emph{orbital} angular momentum eigenstates \cite{Allen1992} respectively. Each point on the Poincar\'e sphere, both on the polarization and the spatial-mode one, may be put in one-to-one correspondence with a unit three-dimensional real vector $\vec{S} = \{S_1/S_0, S_2/S_0, S_3/S_0\}$, where $S_0^2 = S_1^2 + S_2^2 + S_3^2$, as shown in Fig. \ref{fig:poincare}. The three components $S_i$, $i \in \{1, 2, 3\}$ are known as the Stokes parameters \cite{AzzamBook} and $S_0$ denotes the total intensity of the beam.

In this section, we present a novel representation for cylindrically polarized optical beams based on the simultaneous use of both the PPS and the SMPS representation. We show that it is possible, under certain assumptions, to connect the two polarization and spatial representations to obtain two \emph{hybrid} Poincar\'e spheres (HPSs) that embody all characteristics of these complex modes of the electromagnetic field. Up to now, only \emph{uniform} polarization patterns could have been displayed as points on the PPS. However, with this new representation, \emph{non-uniform} polarization patterns are represented for the first time. This is especially helpful -- let it be in theoretical considerations or in experiments -- if one investigates effects deriving from these spatially dependent, complex polarization patterns. Furthermore, manipulations of states within this representation in accordance with the conventional tools such as the PPS and SMPS can be carried out. This is investigated in Sec. 5.

In principle, the two-dimensional polarization and spatial spaces, respectively spanned by the bases $\{ \hat{\mathbf{x}}, \hat{\mathbf{y}} \}$ and $\{ \psi_{10}, \psi_{01}\}$, may be combined in several different manners to produce a four-dimensional mixed polarization-spatial mode space. A \emph{complete} representation of such a space would require the simultaneous display of eight real numbers which is impossible in a three dimensional space. Consequently, our aim is to find an appropriate way for describing this space by using conventional tools, such as the PPS or the SMPS respectively. We begin by noticing that if we are permitted to discard some information, then a lower-dimension visual representation may be possible. The characteristics of this picture would then be fixed by the kind of degree of freedom (DOF) we choose to represent. Driven by certain recent experimental results \cite{Gabriel2010}, here we look for a \emph{hybrid} representation, which would be suitable to describe physical situations where co- and counter-rotating modes do not mix with each other. This absence of mixing occurs, for example, when the light propagates through cylindrically symmetric optical systems as fibres, non-astigmatic lenses, et cetera. In this case, the disregarded information would amount to the relative phase and amplitude between the sets of co- and counterrotating modes. The representation deriving from these constraints is illustrated in Fig. \ref{fig:kombination}.  
The North and South Pole of sphere (c) represent radially and azimuthally polarized states whose total angular momentum (orbital plus spin) is equal to zero. Analogously, sphere (d) represent counter-radial and counter-azimuthal states, still with total zero angular momentum. Each of these two pairs of states represent a novel kind of \emph{hybrid} DOF of the electromagnetic field, since such states describe neither purely polarized nor purely spatial modes of the field. 

Mathematically, the combination of the SMPS and the PPS is realized by forming the four dimensional basis given by the Cartesian product of the the spatial mode $\{  \psi_{10}, \psi_{01}  \}$  and polarization $\{ \mathbf{\hat x}, \mathbf{\hat y}\}$ bases: 
\begin{equation}
\{  \psi_{10}, \psi_{01}  \} \otimes \{ \mathbf{\hat x}, \mathbf{\hat y}\}=\{  \psi_{10}\mathbf{\hat x},  \psi_{10}\mathbf{\hat y}, \psi_{01}\mathbf{\hat x}, \psi_{01}\mathbf{\hat y} \}. 
\end{equation}
To span this four dimensional space, we choose the basis $\{\hat{\mathbf{u}}_R^{+}, \hat{\mathbf{u}}_A^{+}, \hat{\mathbf{u}}_R^{-},  \hat{\mathbf{u}}_A^{-}  \}$. As it was shown above, these two sets of vector functions, namely the ``$+$'' and ``$-$'' ones, do not mix under rotation and, moreover, obey the two fundamental rotation laws in Eqs. (\ref{rot}) and (\ref{rotMinus}). Such a physical constraint may be mathematically implemented by a \emph{superselection rule} \cite{Peres} which forbids interference between states separated according to the law in Eq. (\ref{rot}). Therefore, we use this law to split the four dimensional space $\{  \psi_{10}\mathbf{\hat x},  \psi_{10}\mathbf{\hat y}, \psi_{01}\mathbf{\hat x}, \psi_{01}\mathbf{\hat y} \} $ into the Cartesian sum of two subspaces spanned by the ``$+$'' and ``$-$'' sets of modes:
\begin{equation}
 \{  \psi_{10}, \psi_{01}  \} \otimes \{ \mathbf{\hat x}, \mathbf{\hat y}\}= \{\mathbf{u}_R^+,\mathbf{u}_A^+\} \oplus \{\mathbf{u}_R^-,\mathbf{u}_A^-\}.
\end{equation}
This shows, that depending upon their behavior under rotation, cylindrically polarized optical beams may be subdivided in two independent sets which can be represented by points on the surface of two distinct hybrid Poincar\'e spheres, as illustrated in Fig. \ref{fig:kombination}. 
It is worth to stress once again that a beam represented by an arbitrary superposition of either the standard basis vectors $\{  \psi_{10}\mathbf{\hat x},  \psi_{10}\mathbf{\hat y}, \psi_{01}\mathbf{\hat x}, \psi_{01}\mathbf{\hat y} \}$ or the cylindrical basis vectors $\{\mathbf{u}^+_R, \mathbf{u}^+_A, \mathbf{u}^-_R, \mathbf{u}^-_A\}$ needs four complex numbers to be described completely. This amounts to eight real numbers which can be reduced to seven because of normalization. 
However, our superselection rule defined above reduces the number of real parameters necessary to describe cylindrically polarized beams further to $3\oplus3$, thus permitting the introduction of a ``two-Poincar\'e Cartesian sphere'' representation.
\begin{figure}
\centering
\includegraphics[width=10cm]{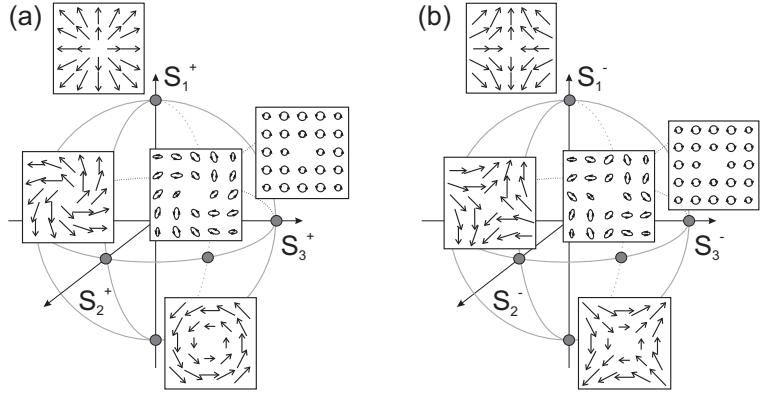}
\caption{Polarization states on the HPSs, represented on the sphere of the ``$+$'' modes (a) and the ``$-$'' modes (b).}
\label{fig:hauptpunkte}
\end{figure}

Besides evident visualization properties, our representation permits to describe straight forwardly cylindrically polarized states of light, in terms of \emph{hybrid} Stokes parameters, as opposed to the either polarization or spatial Stokes parameters. These hybrid Stokes parameters convey information about both polarization and spatial-mode degrees of freedom simultaneously. They are naturally defined as: 
\begin{subequations}
\begin{align}
S_0 ^{\pm}&=  {f_R^{\pm}}^* f_R^{\pm} + {f_A^{\pm}}^* f_A^{\pm}, \\
S_1^{\pm} &= {f_R^{\pm}}^* f_R^{\pm} - {f_A^{\pm}}^* f_A ^{\pm}, \\
S_2^{\pm} &=  {f_R^{\pm}}^* f_A^{\pm} +{ f_A^{\pm}}^* f_R ^{\pm}, \\
S_3^{\pm} &=-i(   {f_R^{\pm}}^* f_A^{\pm}-  {f_A^{\pm}}^* f_R^{\pm}),
\end{align}
\end{subequations}
where the symbol $\/^*$ denotes the complex conjugation. $f_A^{\pm} = (\mathbf{u}^{\pm}_A, \mathbf{E})$ and $ f_R^{\pm} = (\mathbf{u}^{\pm}_R, \mathbf{E})$ are the field amplitudes of the electric field vector in the bases $\{ \mathbf{u}^{\pm}_A, \mathbf{u}^{\pm}_R \}$, where 
\begin{equation}
\mathbf{E} = f_A^{\pm} \mathbf{u}^{\pm}_A+ f_R^{\pm} \mathbf{u}^{\pm}_R.
\end{equation}
These amplitudes can be expressed in terms of the spherical coordinates on the two independent hybrid Poincar\'e spheres (HPSs) as 
\begin{subequations}
\begin{align}
&f_A ^{\pm}= \cos{(\theta/2)}, \\
&f_R^{\pm} =   \exp(i\phi)\sin{(\theta/2)}. 
\end{align}
\end{subequations}
The first sphere of the HPSs represents the ``$+$'' modes [Fig. \ref{fig:hauptpunkte} (a)] and the second the ``$-$'' [Fig. \ref{fig:hauptpunkte} (b)] modes. It is possible to show that these Stokes parameters can be actually measured by means of conventional optical elements. This characteristic is particularly relevant for the possible quantum applications of our formalism, where simultaneous measurability of Stokes parameters describing spatially separated optical beams, is crucial \cite{Korolkova2002}.

Concerning the structure of the HPSs, an interesting feature is that every point on the meridian between the North and the South Pole ($\theta^{\pm}\in[ 0 , \pi] $, $\phi^{\pm} = 0$) describes a locally linear polarization state as illustrated in both Figs. \ref{fig:hauptpunkte} (a) and (b). Except for these points and the two ones on the $S_3$ axis ($ \theta = \frac{\pi}{2}, \phi = \{\frac{\pi}{2},\frac{3}{2}\pi \}$) which represent circularly polarized states, all remaining points on the spheres describe states of non-uniform elliptical polarization. This is fully consistent with the structure of the conventional PPS or SMPS representation.

\section{Manipulating states on the HPSs}
\begin{figure}
\centering
\includegraphics[width=12cm]{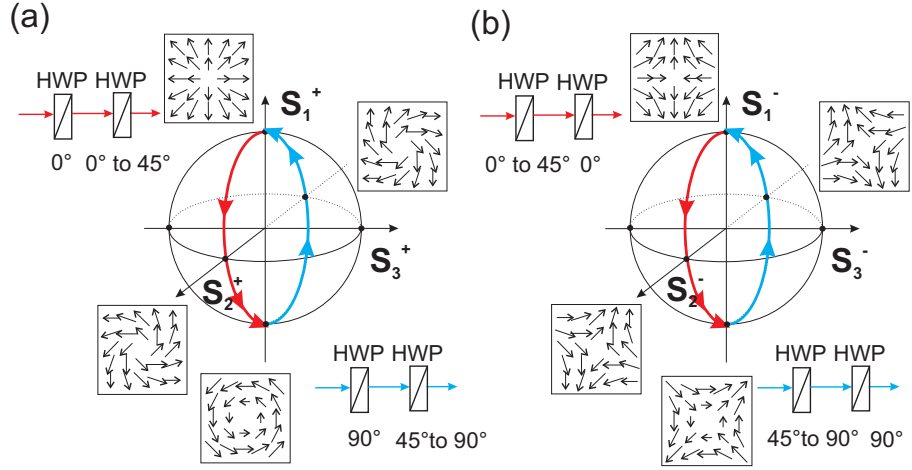}
\caption{Manipulation of states on the HPSs, represented on the sphere of the ``$+$'' modes (a) and the ``$-$'' modes (b).}
\label{fig:trajektorien}
\end{figure}
As in terms of the conventional PPS and SMPS representation, conversions of states by means of polarization and spatial phase retarders such as waveplates or cylindrical lenses, respectively, are possible. In general, waveplates only act on the polarization part of the vector beam and cylindrical lenses on the spatial mode part, as it was shown in \cite{Allen1992}. In terms of the PPS, two independently adjustable quaterwaveplates (QWPs) and one halfwaveplate (HWP) are sufficient to control polarization \cite{AzzamBook}, so reaching every point on this sphere is possible. In analogy to this, manipulating states on the SMPS is possible, as well. However, one cannot use waveplates but instead cylindrical lenses \cite{Padgett1999} to control the spatial mode. Displaying the information from both the PPS and the SMPS simultaneously with the new representation, combinations of both waveplates and cylindrical lenses are necessary. The process of converting a state on the HPSs to a different one -- for instance, going from $S_1^{\pm}$ via $S_2^{\pm}$, $-S_1^{\pm}$ and  $-S_2^{\pm}$ back to $S_1^{\pm}$ -- can be achieved by applying two halfwaveplates and rotating one of these as illustrated in Fig. \ref{fig:trajektorien}. The fixed HWP induces a local flip of the vectors in the polarization pattern and the other one rotates continuously these vectors. However, due to the special cylindrical symmetry of our states, no ``arbitrary-to-arbitrary'' \cite{DamaskBook} transformations with the help of waveplates or cylindrical lenses are permitted. Therefore, only a few transformations can be accomplished without altering the symmetry of the state. These are listed below:
\begin{subequations}
\begin{align}
\{ \theta, \phi\} \rightarrow &\{\theta ' = \pi - \theta, \phi \} &\phi \in \{ 0, \frac{\pi}{2}, \pi,\frac{3\pi}{2} \},\\
\{ \theta, \phi\} \rightarrow &\{\theta ' = \pi - \theta, \phi ' = \pi +\phi\} & \phi \in \{ 0, \frac{\pi}{2}, \pi,\frac{3\pi}{2} \},\\
\{ \theta, \phi\} \rightarrow &\{\theta ' = \theta, \phi' = \pi +\phi  \} &\phi \in [0, 2\pi[,
\end{align} 
\end{subequations}
\begin{figure}[!ht]
\centering
\includegraphics[width=8.5cm]{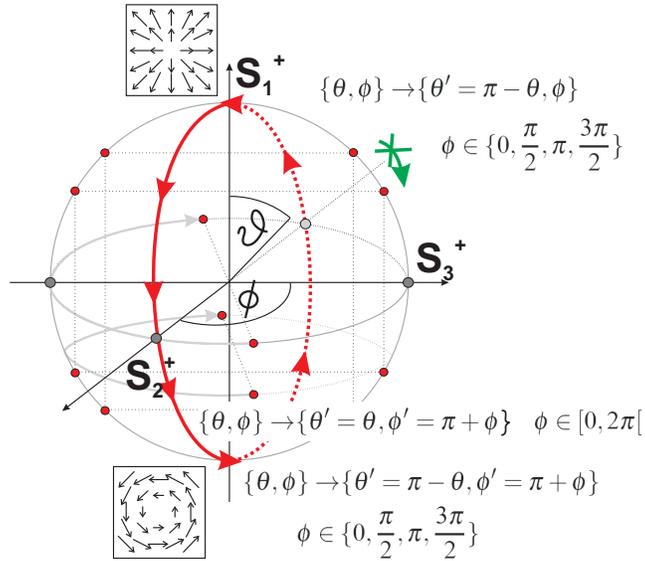}
\caption{Transformations on the ``+''  HPS: Without altering the rotational symmetry of the state, these transformations are possible with the help of \emph{conventional} phase retarders.} 
\label{fig:Zugelassene_Transformationen}
\end{figure}
where $\theta \in [0,\pi]$ and $\phi$ refers to the angles parametrizing a sphere, as shown in Fig. \ref{fig:Zugelassene_Transformationen}. These constraints are due to the fact that phase retarders imprint a spatially uniform phase shift onto the cross section of the beam, which is not, per definition, cylindrically invariant. These constraints are explicitly deduced in Appendix. 

\begin{figure}[!ht]
\centering
\includegraphics[width=10cm]{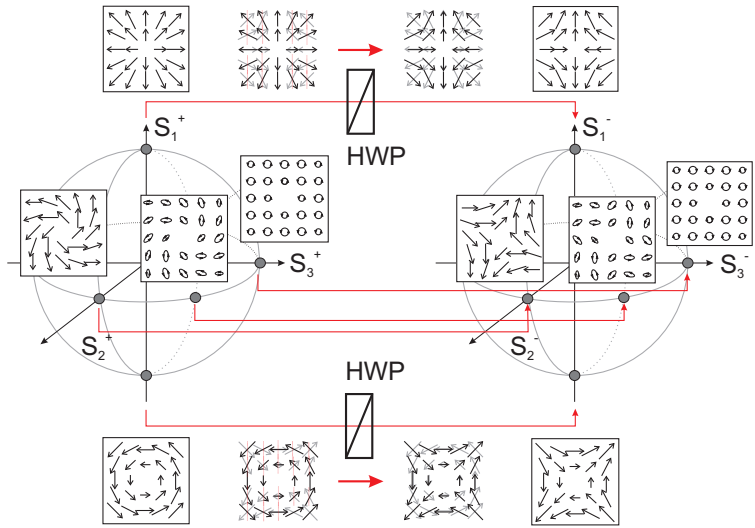}
\caption{By a mirror image symmetry operation one can pass from one sphere to the respective other sphere.} 
\label{fig:mirror_image}
\end{figure}

To complete the discussion about possible manipulations, we notice that the two sets $\{\mathbf{u}_R^+,\mathbf{u}_A^+\}$ and $\{\mathbf{u}_R^-,\mathbf{u}_A^-\}$ are connected to each other by a mirror image symmetry operation, which is illustrated in Fig. \ref{fig:mirror_image}. This can be practically performed by a halfwaveplate \cite{DamaskBook}. 
So, jumps between the ``$+$'' and the ``$- $'' spheres can be carried out easily and they considerably enlarge the number of possible transformation. 
\section{Quantum properties of cylindrically polarized states of light}

As we have discussed only \emph{classical} properties of cylindrically polarized states of light so far, we now cover their quantum aspects. In this section, we give a few examples of application of our hybrid polarization/spatial formalism to some quantum states of light.

To begin with, let us write the electric-field \emph{quantum operator} for paraxial beams of light in the same units of Eq. (\ref{Efeld}) as:
\begin{equation}\label{QEfield}
 \hat{\mathbf{E}}(\mathbf{r}, t) =   \hat{\mathbf{E}}^+(\mathbf{r}, t) + \hat{\mathbf{E}}^-(\mathbf{r}, t),
\end{equation}
where $\hat{\mathbf{E}}^- = (\hat{\mathbf{E}}^+)^\dagger$, and
\begin{equation}\label{QEfieldPlus}
\hat{\mathbf{E}}^+(\mathbf{r}, t) =  \frac{1}{2}\sum_{\lambda =1}^2 \sum_{n,m} \hat{\mathbf{e}}_\lambda \hat{a}_{\lambda n m} \psi_{nm} e^{- i \chi},
\end{equation}
with $\hat{\mathbf{e}}_1 = \hat{\mathbf{x}}$ and $\hat{\mathbf{e}}_2 = \hat{\mathbf{y}}$ and $n,m \in \{0,1,2,\ldots \}$.
The paraxial annihilation operators $\hat{a}_{\lambda n m}$ satisfy the canonical commutation relations
\begin{equation}\label{CCR}
\big[ \hat{a}_{\lambda n m},\hat{a}_{\lambda' n' m'}^\dagger\big] = \delta_{\lambda \lambda'}
\delta_{n n'}\delta_{m m'}.
\end{equation}

In order to have a correct description of quantum noise in light beams with cylindrical polarization, we need to establish first a correspondence between the classical and the quantum representation of these beams.  For sake of simplicity, here we restrict our attention to co-rotating modes solely, namely $\mathbf{u}_A^+$ and $\mathbf{u}_R^+$.
Having in mind this goal, let $\mathbf{u}_{AB}(\mathbf{r})$ be the generic cylindrical mode (the superscript ``$+$'' will be omitted from now on) that now we write as
\begin{align}\label{82}
\mathbf{u}_{AB}(\mathbf{r}) = & \; \frac{1}{\sqrt{2}} \left[ \hat{\mathbf{x}} \left( A \psi_{10}  - B \psi_{01} \right) +  \hat{\mathbf{y}} \left( B \psi_{10}  + A \psi_{01} \right) \right] \nonumber \\
= & \;  A \mathbf{u}_R + B \mathbf{u}_A ,
\end{align}
where Eqs. (\ref{eq:uR+uA+a}) and (\ref{eq:uR+uA+b}) were used in the second line and again
 $|A|^2 + |B|^2 = 1$ is assumed.
 Since $\left( \mathbf{u}_R, \mathbf{u}_A \right)=0$, then two different modes $\mathbf{u}_{AB}(\mathbf{r})$ and $\mathbf{u}_{A'B'}(\mathbf{r})$ satisfy the relation
\begin{align}\label{83}
\left(\mathbf{u}_{AB}(\mathbf{r}) , \mathbf{u}_{A'B'}(\mathbf{r}) \right) = A^* A' + B^* B'.
\end{align}
 Now the question is: What is the correct form of the field operator $\hat{a}_{AB}$ that annihilates a photon in the  mode $\mathbf{u}_{AB}(\mathbf{r})$? We seek an answer to this question by requiring that:
\begin{enumerate}
\item The \emph{coherent} state $| \alpha \rangle = \exp(\alpha \hat{a}_{AB}^\dagger -  \alpha^* \hat{a}_{AB}) \lvert 0 \rangle$ must produce a  coherent signal $\mathcal{S} = \langle \alpha | \hat{\mathbf{E}}| \alpha \rangle$  equal to the classical electric field:
\begin{align}\label{84}
\mathcal{S} =
\frac{1}{2} \left[ \alpha \, \mathbf{u}_{AB}(\mathbf{r}) e^{- i \chi}  + \alpha^* \mathbf{u}^*_{AB}(\mathbf{r}) e^{ i \chi} \right].
\end{align}
\item The \emph{single-photon} state $|1 \rangle = \hat{a}^\dagger_{AB} | 0 \rangle$ must generate a  photon wave-function $\langle 0 | \hat{\mathbf{E}}^+ | 1 \rangle $  equal to:
\begin{align}\label{Azi275}
\langle 0 | \hat{\mathbf{E}}^+ | 1 \rangle = \frac{1}{2}  \mathbf{u}_{AB}(\mathbf{r}) e^{- i \chi}.
\end{align}
\end{enumerate}
With the help of Eq. (\ref{82}) we guess the following form for the sought annihilation operator $\hat{a}_{AB}$:
\begin{align}\label{86}
\hat{a}_{AB} = A \left(  \frac{\hat{a}_{x10} + \hat{a}_{y01}}{\sqrt{2}} \right) + B \left(\frac{-\hat{a}_{x01} + \hat{a}_{y10}}{\sqrt{2}} \right),
\end{align}
where we have used the polarization indexes $x$ for $\lambda=1$ and $y$ for $\lambda=2$.
From Eqs. (\ref{CCR},\ref{86}) it follows that:
\begin{align}\label{Azi290}
 \big[ \hat{a}_{A'B'}, \hat{a}_{AB}^\dagger \big] = A^* A' + B^* B',
\end{align}
which, together with Eq. (\ref{83}), shows that $\hat{a}_{A'B'}$ and $\hat{a}_{AB}^\dagger$ commute whenever the two modes $\mathbf{u}_{AB}(\mathbf{r})$ and $\mathbf{u}_{A'B'}(\mathbf{r})$ are orthogonal.
\subsection{Coherent states}
Let $\hat{D}_i (\beta), \; i \in \{1,2,\ldots, \infty \}$ be the \emph{coherent-state displacement operator} defined as
\begin{equation}
\hat{D}_i(\beta) = \exp \big( \beta \hat{a}_i - \beta^* \hat{a}_i^\dagger \big),
 \end{equation}
where the single label $i$ embodies $\lambda,n,m$, so that $\sum_i = \sum_\lambda \sum_n \sum_m$, and we choose the first $4$ values of the index $i$ in such a way that
\begin{equation}\label{label}
 \hat{a}_1 = \hat{a}_{x10}, \quad  \hat{a}_2 = \hat{a}_{y01}, \quad  \hat{a}_3 = \hat{a}_{x01}, \quad  \hat{a}_4 = \hat{a}_{y10}.
 \end{equation}
If we rewrite the field mode functions as
\begin{equation}
\bm{v}_i(\mathbf{r}, t) \equiv   \hat{\mathbf{e}}_\lambda  \psi_{nm} \exp(- i \chi),
\end{equation}
then Eq. (\ref{QEfieldPlus}) takes the simpler form
\begin{equation}\label{QEfieldPlus2}
\hat{\mathbf{E}}^+ =  \frac{1}{2}\sum_{i =1}^\infty\bm{v}_i \hat{a}_i  .
\end{equation}
Since any coherent state $ | \beta \rangle_i \equiv   \hat{D}_i(\beta) | 0 \rangle $
 satisfies the relation
\begin{equation}\label{QEfieldPlus3}
\hat{\mathbf{E}}^+ | \beta \rangle_i = \frac{1}{2}\sum_{j =1}^\infty \bm{v}_j  \hat{a}_j | \beta \rangle_i = \frac{1}{2} \, \bm{v}_i \beta   | \beta \rangle_i,
\end{equation}
it immediately follows that if we define $ | \alpha_1, \alpha_2 , \alpha_3 , \alpha_4 \rangle \equiv \hat{D}_4(\alpha_4) \hat{D}_3(\alpha_3) \hat{D}_2(\alpha_2) \hat{D}_1(\alpha_1) \lvert 0 \rangle $, then
\begin{align}
\hat{\mathbf{E}}^+ | \alpha_1, \alpha_2 , \alpha_3 , \alpha_4 \rangle = & \, \frac{1}{2}\sum_{i =1}^\infty \bm{v}_i \hat{a}_i  | \alpha_1, \alpha_2 , \alpha_3 , \alpha_4 \rangle
\nonumber \\
= & \, \frac{1}{2} \sum_{i =1}^4 \bm{v}_i  \alpha_i  | \alpha_1, \alpha_2 , \alpha_3 , \alpha_4 \rangle,
 \label{QEfieldPlus4}
\end{align}
which trivially implies that
\begin{align}
\mathcal{S} = & \,  \langle  \alpha_1, \alpha_2 , \alpha_3 , \alpha_4 \lvert \hat{\mathbf{E}}^+ +  \hat{\mathbf{E}}^-  | \alpha_1, \alpha_2 , \alpha_3 , \alpha_4 \rangle
\nonumber \\
= & \, \frac{1}{2} \sum_{i =1}^4 \left(  \bm{v}_i  \alpha_i +  \bm{v}_i^*  \alpha_i^* \right) \nonumber \\
= & \,  \frac{1}{2} \left[ \hat{\mathbf{x}} \left( \alpha_1 \psi_{10} + \alpha_3 \psi_{01} \right) +  \hat{\mathbf{y}} \left( \alpha_4 \psi_{10}  + \alpha_2 \psi_{01} \right) \right]e^{- i \chi} \nonumber \\
 & \quad + \text{c.c.}\label{QEfieldPlus5}
\end{align}
Comparison of Eq. (\ref{QEfieldPlus5}) with Eq. (\ref{82}) shows that if we choose
\begin{equation}\label{Cond28}
\alpha_1 = \alpha_2 = A^* \frac{\alpha}{\sqrt{2}}, \qquad \alpha_3 = -\alpha_4  = -B^* \frac{\alpha}{\sqrt{2}},
\end{equation}
then we have:
\begin{align}\label{Azi300}
 \hat{D}_4(\alpha_4)\hat{D}_3(\alpha_3)  & \hat{D}_2(\alpha_2) \hat{D}_1(\alpha_1) \nonumber \\
& \,=  \exp\left[ \sum_{i=1}^4 \bigl(  \alpha_i  \hat{a}_i^\dagger +  \alpha_i^*  \hat{a}_i \bigr) \right]\nonumber \\
& \,=
\exp \bigl (\alpha \hat{a}_{AB}^\dagger -  \alpha^* \hat{a}_{AB} \bigr),
\end{align}
and Eq. (\ref{84}) is automatically satisfied.
\subsection{Fock states}
The validity of Eq. (\ref{Azi275}) can be checked by defining the  single-photon ladder operator $\hat{A}(\xi)$ as:
\begin{equation}\label{SinglePhotOp1}
\hat{A}(\xi)= \sum_{i=1}^4 \xi^*_i \hat{a}_i, \qquad \text{with} \qquad \sum_{i=1}^4 |\xi_i|^2 =1,
\end{equation}
that satisfies the canonical commutation relations 
 \[
     [\hat{A}(\xi),\hat{A}^\dagger(\xi)]=1. 
 \]
 Let $|1\rangle$ be the single-photon state defined as $|1\rangle = \hat{A}^\dagger(\xi)|0 \rangle$.
Then
\begin{align}
\langle 0 | \hat{\mathbf{E}}^+ |1\rangle = & \,  \frac{1}{2}\langle 0 | \sum_{i=1}^\infty \bm{v}_i \hat{a}_i \sum_{j=1}^4 \xi_j \hat{a}_j^\dagger  |0 \rangle
\nonumber \\
= & \,  \frac{1}{2} \sum_{i,j}  \xi_j \bm{v}_i \langle 0 | \hat{a}_i  \hat{a}_j^\dagger  |0 \rangle
\nonumber \\
= & \,  \frac{1}{2} \sum_{i=1}^4 \bm{v}_i  \xi_i.
\label{QEfieldPlus6}
\end{align}
Now,  if we choose
\begin{equation}\label{Cond29}
\xi_1 = \xi_2 = A^* \frac{1}{\sqrt{2}}, \qquad \xi_3 = -\xi_4  = -B^* \frac{1}{\sqrt{2}},
\end{equation}
then we obtain
\begin{align}
\hat{A}(\xi) = & \, A \left(  \frac{\hat{a}_{x10} + \hat{a}_{y01}}{\sqrt{2}} \right) + B \left(\frac{-\hat{a}_{x01} + \hat{a}_{y10}}{\sqrt{2}} \right)\nonumber \\
= & \, \hat{a}_{AB} , \label{Azi310}
\end{align}
in agreement with Eq. (\ref{Azi275}). Thus, we have proved that Eq. (\ref{86}) furnishes the correct expression for  $ \hat{a}_{AB}$.
\subsection{Squeezed states}
As a last example, here we write explicitly the expression for an azimuthally polarized squeezed state. The extension to the other cylindrically polarized states is straightforward and, therefore, will be omitted. As a result of such calculation, we shall find that \emph{classical} inseparability automatically lead to \emph{quantum} entanglement when the azimuthally polarized beam is prepared in a nonclassical state \cite{Gabriel2010}.

The ``natural'' definition of \emph{squeezing operator} for the azimuthally polarized field is
\begin{align}\label{Sr}
\hat{S}_A(\zeta) = \exp \Big[\frac{1}{2} \zeta^* \big( \hat{a}_A \big)^2 -  \frac{1}{2}\zeta \big( \hat{a}_A^\dagger \big)^2  \Big] ,
\end{align}
where $\zeta = s e^{i \vartheta}$ is the complex squeezing parameter. By substituting Eq. (\ref{86}) evaluated for $A=0, \, B=1$  into Eq. (\ref{Sr}), it is not difficult to obtain
\begin{align}\label{Sr2}
\hat{S}_A(\zeta) = \hat{S}_3(\zeta/2) \hat{S}_4(\zeta/2) \hat{S}_{34}(-\zeta/2),
\end{align}
where we have defined the one- and two-mode squeezing operator
\begin{align}
\hat{S}_i(\zeta) = & \, \exp \Big[ \frac{1}{2}\zeta^* \big( \hat{a}_i \big)^2 - \frac{1}{2}\zeta \big( \hat{a}_i^\dagger \big)^2  \Big], \label{Si} \\
 \hat{S}_{ij}(\zeta) = & \, \exp \Big( \zeta^* \hat{a}_i \hat{a}_j - \zeta \hat{a}_i^\dagger \hat{a}_j^\dagger  \Big).\label{S12}
\end{align}
Thus, the squeezed-vacuum state
\begin{align}\label{Sq1}
|\zeta \rangle_A = \hat{S}_3(\zeta/2) \hat{S}_4(\zeta/2) \hat{S}_{34}(\zeta)| 0 \rangle,
\end{align}
is clearly entangled, since
\begin{align}\label{Sq2}
\hat{S}_{34}(\zeta)| 0 \rangle  =  \frac{1}{\cosh s} \sum_{n=0}^\infty (e^{-i \vartheta}\tanh s)^n |n \rangle_3 |n \rangle_4,
\end{align}
represents the well-known entangled two-mode squeezed vacuum \cite{Bowen2003}. 
Thus, we have shown that classical inseparability naturally lead to quantum entanglement.

%
%
\section{Conclusion}
In conclusion, we have established a proper theoretical framework for a classical description of cylindrically polarized states of the electromagnetic field. We have highlighted the rotation principles of these modes which can be derived from their very particular intrinsic symmetries. Moreover, we found that all CPMs have a zero total angular momentum and are clearly non-separable which can be proven by their Schmidt rank. This subtlety leads to intriguing features of these modes in both the classical and quantum domain of optics, the latter being recently demonstrated in experiments \cite{Gabriel2010}. We have exploited these findings to present a new way of visualizing CPMs on a pair of hybrid Poincar\'e spheres. This permits, at a glance, to easily recognize how phase retarders such as waveplates or cylindrical lenses  act on these complex polarization patterns. Furthermore, we have extended our discussion of the properties of these modes to the quantum domain and have shown that the classical inseparability property leads to the intriguing feature of entanglement.

In general, our results presented here give insight into the theoretical structure of these modes and their particular behaviour. We believe that the presented formalism and results are of utility to both the quantum and the classical communities especially in view of the recently growing interest in the theory and applications of light beams with complex spatial and polarization patterns.

The authors thank Peter Banzer for fruitful discussions.


\section{Appendix}
In our hybrid Poincar\'e sphere representation, only certain transformations are permitted. Due to the fact that these constraints are not intuitive, we demonstrate here how they can be derived. Generally, besides the particular cases in illustrated Fig. \ref{fig:Zugelassene_Transformationen}, the fundamental rules that an optical transformation must obey in order to not violate the global rotation law in Eq. (\ref{rot}) and consequently not to generate mixing between the co- and counter-rotating modes, can be deduced from perfectly general principles, as follows.

Consider a generic transformation $\mathscr{T} = \mathscr{M} \times \mathscr{P}$ that acts upon both spatial ($\mathscr{M}$) and polarization ($\mathscr{P}$) degrees of freedom of the field separately. Here, it is convenient to adopt a quantum-like  notation, and write
\begin{subequations}\label{Azi150}
\begin{align}
\hat{\mathbf{x}} &\doteq  | e_1 \rangle,  \quad & \hat{\mathbf{y}} & \doteq  | e_2 \rangle , \label{Azi150a} \\
\psi_{10}(\mathbf{x}) & \doteq  | \psi_1(\mathbf{x}) \rangle,  \quad & \psi_{01}(\mathbf{x}) & \doteq  | \psi_2(\mathbf{x}) \rangle, \label{Azi150b}
\end{align}
\end{subequations}
where the symbol ``$\doteq$'' stands for: ``is represented by''. Note, that the quantum-mechanical notation is just a convenient manner of writing our \emph{classical} modes of  the electro-magnetic field and it has nothing to do with \emph{quantum-mechanical interpretation}. However, for the sake of completeness, our hybrid classical description will be extended to the quantum domain in Appendix II.
Thus, for example, our set of cylindrically polarized modes may be represented as:
\begin{align}\label{Azi160}
\mathbf{u}_R^\pm \doteq | u_R^\pm \rangle = \frac{1}{\sqrt{2}} \left( \pm | \psi_1 \rangle| e_1 \rangle + | \psi_2 \rangle| e_2 \rangle \right),  \\
\mathbf{u}_A^\pm \doteq| u_A^\pm \rangle = \frac{1}{\sqrt{2}} \left( \mp | \psi_2 \rangle| e_1 \rangle + | \psi_1 \rangle| e_2 \rangle \right).
\end{align}
In passing, we note that these states have the formal structure of the quantum Bell-states. More specifically, if one establishes the following formal equivalence between the two qubits $(A,B)$ Hilbert space and our two degrees of freedom (polarization and spatial mode) space: $|0_A\rangle\sim \mathbf{\hat x},\, |1_A\rangle \sim \mathbf{\hat y}$, $|0_B\rangle\sim \psi_{10},\, |1_B\rangle \sim \psi_{01}$, then it is not difficult to show that our basis is mathematically equivalent to the quantum Bell basis \cite{Nielsen}:
\begin{equation}
\{ \mathbf{u}_R ^+, \,\mathbf{u}_A ^+,\mathbf{u}_R ^-, \,\mathbf{u}_A ^- \} \sim \{ |\Phi ^+ \rangle,  -|\Psi^- \rangle, - |\Phi ^- \rangle,  |\Psi ^+ \rangle    \},
\end{equation} 
where $ |\Phi ^\pm \rangle = \left (|0_A,0_B\rangle \pm |1_A,1_B\rangle\right)/\sqrt{2}$ and $|\Psi ^\pm \rangle = \left (|0_A,1_B\rangle \pm |1_A,0_B\rangle\right)/\sqrt{2}$. This analogy has been recently exploited in \cite{Borges2010}. 

Now, let us rewrite Eqs. (\ref{Azi140a}) and (\ref{Azi140b}) in quantum-like notation as:
\begin{align}\label{Azi170}
| u^\pm (\mathbf{x})\rangle = & \; \hat{G}^\pm(\varphi)| u^\pm (\mathbf{x})\rangle ,
\end{align}
where the two \emph{operators} $\hat{G}^\pm(\varphi)$ are determined by  their action upon the generic basis state $| \psi_i (\mathbf{x})\rangle | e_j \rangle$. According to Eqs. (\ref{Azi130}) one has:
\begin{align}\label{Azi180}
\hat{G}^\pm(\varphi)| \psi_i (\mathbf{x})\rangle | e_j \rangle = & \; | \psi_i (R(\mp \varphi)\mathbf{x})\rangle \hat{R}(\varphi)| e_j \rangle ,
\end{align}
where the rotation operator $\hat{R}(\varphi)$ is expressed in terms of the matrix elements ${R}_{kj}(\varphi)$ of the rotation matrix $R(\varphi)$ as:
\begin{align}\label{Azi190}
\hat{R}(\varphi) | e_j \rangle = \sum_{k=1}^2 {R}_{kj}(\varphi)| e_k \rangle,
\end{align}
From Eqs. (\ref{HermiteGauss_a}) and (\ref{HermiteGauss_b}) it follows that $\psi_{10}$ and $\psi_{01}$ transform as the components of the position vector $\mathbf{x}$, namely:
\begin{subequations}\label{Azi210}
\begin{align}
\psi_{10} (R(\mp \varphi)\mathbf{x}) = & \; \psi_{10} (\mathbf{x}) \cos \varphi \pm \psi_{01} (\mathbf{x}) \sin \varphi, \label{Azi210a} \\
\psi_{01} (R(\mp \varphi)\mathbf{x}) = & \; \mp \psi_{10} (\mathbf{x}) \sin \varphi + \psi_{01} (\mathbf{x}) \cos \varphi. \label{Azi210b}
\end{align}
\end{subequations}
These formulas may be put straightforwardly in the operator form which reads:
\begin{align}\label{Append220}
| \psi_i (R(\mp \varphi)\mathbf{x})\rangle = & \; \hat{R}(\mp \varphi) | \psi_i (\mathbf{x})\rangle \nonumber \\
 = & \;  \sum_{l=1}^2 {R}_{il}( \mp \varphi)| \psi_l (\mathbf{x})\rangle.
\end{align}
Note the different order of the indices in the matrix elements appearing in Eq. (\ref{Azi190}) and Eq. (\ref{Append220}).
Finally, gathering all these formulas together we  obtain:
\begin{align}\label{Azi230}
\hat{G}^\pm(\varphi)| \psi_i (\mathbf{x})\rangle & | e_j \rangle = \hat{R}(\mp \varphi) | \psi_i (\mathbf{x})\rangle \hat{R}(\varphi) | e_j \rangle \nonumber \\
 = & \; \sum_{k, \,l} {R}_{il}( \mp \varphi)| \psi_l (\mathbf{x})\rangle {R}_{kj}(\varphi)| e_k \rangle,
 \nonumber \\
 = & \; \sum_{k, \,l} {R}_{il}( \mp \varphi) {R}_{jk}^T(\varphi)| \psi_l (\mathbf{x})\rangle| e_k \rangle,\nonumber \\
 = & \; \sum_{k, \,l} \big[ R(\mp \varphi) \otimes R(-\varphi) \big]_{ij,lk}| \psi_l (\mathbf{x})\rangle | e_k \rangle,
\nonumber \\
 = & \; \hat{R}(\mp \varphi) \otimes \hat{R}(-\varphi) | \psi_i (\mathbf{x})\rangle | e_j \rangle ,
\end{align}
where $A^T$ indicates the transpose of the generic matrix $A$, and we used the fact that for orthogonal matrices one has $R^T(\varphi)=R(-\varphi)$.
Here, in the direct  product $ \hat{R}(\mp \varphi) \otimes \hat{R}(-\varphi) $  the first operator acts upon the spatial degrees of freedom solely, while the the second one acts on the polarization degrees of freedom.
Since in all previous calculations the basis state $| \psi_i (\mathbf{x})\rangle | e_j \rangle$ was arbitrarily chosen, the relation (\ref{Azi230}) above is perfectly general and may be rewritten as:
\begin{align}\label{Azi240}
\hat{G}^\pm(\varphi)= \hat{R}(\mp \varphi) \otimes \hat{R}(-\varphi) .
\end{align}

Now, let $\hat{M}$ and $\hat{P}$ be such that $\hat{T} = \hat{M} \otimes \hat{P}$, the two spatial and polarization operators corresponding to the \emph{physical} transformation $\mathscr{T} = \mathscr{M} \times \mathscr{P}$. Then, by definition such operation transforms  $| u^\pm (\mathbf{x})\rangle$ into $\hat{T}| u^\pm (\mathbf{x})\rangle$. It is clear that this new state
maintains the same symmetry under rotation of the original state $| u^\pm (\mathbf{x})\rangle$ if and only if:
\begin{align}\label{Azi250}
\hat{G}^\pm(\varphi) \hat{T} | u^\pm (\mathbf{x})\rangle = \hat{T} | u^\pm (\mathbf{x})\rangle.
\end{align}
By multiplying both sides of Eq. (\ref{Azi170}) by $\hat{T}$, one straightforwardly obtains:
\begin{align}\label{Azi255}
\hat{T} | u^\pm (\mathbf{x})\rangle = \hat{T} \hat{G}^\pm(\varphi)| u^\pm (\mathbf{x})\rangle.
\end{align}
 Thus, by using Eq. (\ref{Azi255}) into Eq. (\ref{Append220}) one easily obtain:
\begin{align}\label{Azi260}
\hat{G}^\pm(\varphi) \hat{T} | u^\pm (\mathbf{x})\rangle = & \; \hat{T}\hat{G}^\pm(\varphi)| u^\pm (\mathbf{x})\rangle.
\end{align}
or, equivalently:
\begin{align}\label{Azi270}
 \big[ \hat{G}^\pm(\varphi) \hat{T} -   \hat{T} \hat{G}^\pm(\varphi) \big] |u^\pm (\mathbf{x})\rangle =0.
\end{align}
This equation can be easily recast in commutator form by rewriting it as follows:
\begin{align}\label{Append71}
 \big[ \hat{G}^\pm(\varphi), \hat{T} \big] | u^\pm (\mathbf{x})\rangle =0.
\end{align}
Of course, any operator $\hat{T}$ that commutes with $\hat{G}^\pm(\varphi)$, namely $\big[ \hat{G}^\pm(\varphi), \hat{T} \big]=0$, automatically fulfills Eq. (\ref{Append71}). However, this condition is sufficient but not \emph{necessary} because Eq. (\ref{Append71}) represents a much weaker constraint, as it only requires that the kernel (or, at least, part of it) of the operator $\big[ \hat{G}^\pm(\varphi), \hat{T} \big]$ coincides with the subspace spanned by either $ | u^+ (\mathbf{x})\rangle$ or $ | u^- (\mathbf{x})\rangle$. In other words, all what is required to $\hat{T}$ by Eq. (\ref{Append71}), is that the commutator $\big[ \hat{G}^\pm(\varphi), \hat{T} \big]$ has a null eigenvalue in correspondence of the eigenvector $| u^\pm (\mathbf{x})\rangle$.

Now, having this caveat in mind, we can rewrite the condition (\ref{Append71}) as follows:
\begin{align}\label{Azi280}
\bigl[ \hat{R}(\mp \varphi) \otimes \hat{R}(-\varphi)\bigr] & \bigl[\hat{M} \otimes \hat{P} \bigr] \nonumber \\
& = \bigl[\hat{M} \otimes \hat{P}\bigr]\bigl[\hat{R}(\mp \varphi) \otimes \hat{R}(-\varphi) \bigr] ,
\end{align}
or, equivalently:
\begin{align}\label{Azi285}
 \hat{R}(\mp \varphi)\hat{M} \big[ \hat{R}(\mp  \varphi) \big]^{-1} \otimes \hat{R}(-\varphi) \hat{P} \big[ \hat{R}(-\varphi) \big]^{-1} = \hat{M} \otimes \hat{P},
\end{align}
where all the operators are understood to be restricted to the subspaces spanned by $| g^\pm (\mathbf{x})\rangle$.
Obviously, this commutator expression is satisfied whenever the following two  conditions are separately satisfied:
\begin{align}\label{Append74}
\bigl[ \hat{R}(\varphi) , \hat{M} \bigr] = 0,  \qquad \bigl[ \hat{R}(\varphi) , \hat{P}\bigr]  =0,
\end{align}
where the arbitrariness of the angle $\varphi$ has been exploited. But, again, this condition is sufficient but not at all necessary. In fact, it is not difficult to see that if one choose either 
\begin{align}\label{Append75}
M = \left(
      \begin{array}{cc}
        m_1 & m_2 \\
        -m_2 & m_1 \\
      \end{array}
    \right) \quad \text{and} \quad   P = \left(
      \begin{array}{cc}
        p_1 & p_2 \\
        -p_2 & p_1 \\
      \end{array}
    \right),
\end{align}
or
\begin{align}\label{Append76}
M = \left(
      \begin{array}{cc}
        m_1 & m_2 \\
        m_2 & -m_1 \\
      \end{array}
    \right) \quad \text{and} \quad P = \left(
      \begin{array}{cc}
        p_1 & p_2 \\
        p_2 & -p_1 \\
      \end{array}
    \right),
\end{align}
with $m_1, m_2, p_1,p_2 \in \mathbb{C}$, then Eq. (\ref{Append71}) is always satisfied but Eqs. (\ref{Append74}) are satisfied only by the first choice (\ref{Append75}) that represent orthogonal matrices whenever $m_1^2+m_2^2 = 1 = p_1^2 + p_2^2$.

It is worth to notice that a circular polarizer (left or right) is represented by the Jones matrix
\begin{align}\label{Azi320}
P = \frac{1}{2}\left(
      \begin{array}{cc}
        1 & \pm i \\
        \mp i & 1 \\
      \end{array}
    \right),
\end{align}
which is of the form (\ref{Append75}). Conversely, a half-wave plate with fast axis at angle $\alpha$ with respect to the horizontal axis is represented by the Jones matrix
\begin{align}\label{Azi330}
P = \left(
      \begin{array}{cc}
        \cos(2 \alpha) &  \sin(2 \alpha) \\
         \sin(2 \alpha) &  -\cos(2 \alpha) \\
      \end{array}
    \right),
\end{align}
which is of the form (\ref{Append76}).

\end{document}